\documentclass[letterpaper, 10pt, conference]{ieeeconf}  

\IEEEoverridecommandlockouts                              

\usepackage{graphics} 
\usepackage{epsfig} 
\usepackage{epstopdf}
\usepackage{mathptmx} 
\usepackage{times} 
\usepackage{amsmath} 
\usepackage{amssymb}  
\usepackage{float}
\usepackage{algorithm}
\usepackage{algpseudocode}
\usepackage{multirow}
\usepackage{color}
\usepackage[usenames,dvipsnames]{xcolor}
\renewcommand{\baselinestretch}{0.97}

\title{\LARGE \bf
Nonlinear Stabilization via Control Contraction Metrics: a Pseudospectral Approach for Computing Geodesics 
}

\author{Karen Leung $\qquad$ Ian R. Manchester
\thanks{This work was supported by the Australian Research Council. The authors are with the Australian Centre for Field Robotics
(ACFR), University of Sydney, NSW 2006, Australia
        {\tt\small kleu6124@uni.sydney.edu.au, i.manchester@acfr.usyd.edu.au}}%
}

\begin{document}

\maketitle
\thispagestyle{empty}
\pagestyle{empty}

\begin{abstract}
Real-time nonlinear stabilization techniques are often limited by inefficient or intractable online and/or offline computations, or a lack guarantee for global stability. In this paper, we explore the use of Control Contraction Metrics (CCM) for nonlinear stabilization because it offers tractable offline computations that give formal guarantees for global stability. We provide a method to solve the associated online computation for a CCM controller - a pseudospectral method to find a geodesic. Through a case study of a stiff nonlinear system, we highlight two key benefits: (i) using CCM for nonlinear stabilization and (ii) rapid online computations amenable to real-time implementation. We compare the performance of a CCM controller with other popular feedback control techniques, namely the Linear Quadratic Regulator (LQR) and Nonlinear Model Predictive Control (NMPC). We show that a CCM controller using a pseudospectral approach for online computations is a middle ground between the simplicity of LQR and stability guarantees for NMPC.

\end{abstract}
\section{Introduction}
Control design for a general nonlinear dynamical system continues to be a challenging problem. In general, there are three primary benchmarks when constructing an `ideal' nonlinear controller: (i) a globally stabilizing controller, (ii) tractable offline computations and (iii) fast and tractable online computations amenable for real-time implementation.

A common solution is to linearize the dynamics about a particular operating point or desired trajectory and apply a linear control synthesis techniques, e.g. the Linear Quadratic Regulator (LQR). The offline and online computations are tractable however, the system will in general only be locally stable and may become unstable far from the point of linearization. Thus in general, global stability cannot be guaranteed for LQR. At the other end of the spectrum is Nonlinear Model Predictive Control (NMPC) which solves a finite-horizon optimal control problem (OCP) at each iteration \cite{Allgower}.
Although NMPC generally provides good performance, the associated online optimization problems are generally non-convex due to dynamic constraints, and can be difficult to solve in real time. 

An elegant theory of necessary and sufficient conditions for global stabilizability can be given in terms of control Lyapunov functions (CLF) \cite{sontag_universal_1989}. However, the search for a CLF is generally non-convex, and existing constructive methods such as backstepping rely on the system having a particular triangular structure \cite{Krstic}. In general, unless there is some structure to exploit, the search for a CLF is essentially intractable. 

Contraction theory is an attractive alternative because it combines the simplicity and tractability of linear analysis with formal guarantees of global stability \cite{Lohmiller}. The concept of a Control Contraction Metric (CCM) was introduced in  \cite{CCM, Manchester4} to generalize contraction analysis to constructive control design, and gives sufficient conditions for {\em every} trajectory of a nonlinear system to be stabilizable, a property called {\em universal stabilizability}. The conditions are general enough to be necessary and sufficient for linear systems and feedback linearizable systems. The offline search for a CCM is a convex optimization problem \cite{Manchester4} and for polynomial systems can be solved using sum-of-squares (SOS) programming \cite{Aylward}.

Given a CCM, a state feedback controller can be constructed via an integration along a shortest path (a geodesic) between the system's current state and the target state \cite{CCM, Manchester4}. Therefore the online computation of this scheme requires solving an optimization problem to find a geodesic with respect to the CCM. This is analogous to NMPC; solving an optimization problem at each time step. However note that the CCM approach is generally simpler due to the \textit{lack of dynamic constraints}. Indeed, for many nonlinear systems a state-independent (a.k.a. ``flat'') CCM can be found, for which all geodesics are straight lines \cite{ManchesterTang}, but nontrivial metrics can also be found which can offer better performance.

This paper considers the computation of geodesics in the context of CCM and nonlinear stabilization. In particular, the application of CCM to a stiff nonlinear system which are often difficult to control. Indirect methods to finding a geodesic involves deriving an explicit differential equation (the ``geodesic equation'') via the Euler-Lagrange equation, however for non-trivial metrics this will be difficult to solve \cite{Arnold}. Other methods that have been explored recently include the ``Phase Flow Method'' \cite{Ying}, fast marching \cite{Kimmel}, and graph cuts \cite{boykov_computing_2003}. Although indirect methods would offer greater accuracy and confidence that first-order optimality conditions are met, disadvantages include small radii of convergence and the need to analytically derive the particular necessary conditions for each instance of a problem.

Direct methods involve discretizing a shortest path or optimal control problem, and then solving the resulting nonlinear problem (NLP) using generic methods and this is the approach taken in this paper. Popular methods include multiple shooting (e.g. \cite{diehl_efficient_2009}), direct collocation (e.g. \cite{benson_direct_2006}), and global pseudospectral (e.g. \cite{Elnagar, FahrooRoss, Garg}). 

In this paper, we follow a well studied Chebyshev pseudospectral approach \cite{Williams, FahrooRoss_PS_infRHC, GongFahrooRoss_PS_Cheby, FahrooRoss, ross2012review} to solve the geodesic problem but with modifications to cater for our specific framework. We offer a criteria that describes the accuracy of the geodesic. 
By forming an efficient and rapid solution to the geodesic problem for a CCM, we enable a fast construction of a CCM controller amenable to real-time applications. We will compare the performance of the CCM controller with the LQR and NMPC framework which are often used in real-time applications.

The structure of the paper is as follows: in Section \ref{sec:prelim} we define the problem and recall relevant facts about metrics, geodesics, and contraction analysis; in Section \ref{sec:pseudo} we detail the pseudospectral method applied to the geodesic problem. An example stiff nonlinear system is introduced in Section \ref{sec:exGeo}. In Section \ref{sec:exCon} we present the benefits of using a CCM controller over LQR and NMPC and in Section \ref{sec:finding_a_geodesic} we showcase the efficiency and accuracy of using a pseudospectral method in approximating a geodesic over multiple shooting. Finally we offer some brief conclusions.

\section{Preliminaries and Problem Formulation}\label{sec:prelim}
In this paper we consider nonlinear control affine systems, possibly time dependent, of the form:
\begin{equation}
\dot{x} = f\left(x,t\right) + B(x,t)u
\label{eqn: system dynamics}
\end{equation}
where $x(t) \in \mathbb{R}^n$ is the state, $u(t) \in \mathbb{R}^m$ is the control, and $t\in \mathbb{R}^+ := [0,\,\infty )$. The function $f\,:\, \mathbb{R}^n \times \mathbb{R}^+ \rightarrow \mathbb{R}^n$ is assumed to be smooth, and $B\,:\, \mathbb{R}^n\times\mathbb{R}^+ \rightarrow \mathbb{R}^{n\times m}$ has columns $b_i(x,t)$, $i = 1, 2, ..., m$.

The objective is to design a tractable state-feedback control policy that can globally exponentially stabilize {\em any} feasible trajectory of the system \eqref{eqn: system dynamics}. A system is called {\em universally exponentially stabilizable} if this is possible \cite{Manchester4}.

\subsection{Metrics and Geodesics}

%
%
%
A Riemannian metric equips a smooth manifold -- which for this paper remains $\mathbb R^n$ but can be more general -- with an inner product on each tangent space, i.e  $\delta_x^T M(x)\delta_x$ \cite{Lee}, \cite{boothby_introduction_1986}. This gives a local definition of length and orthogonality. Consider a smooth curve $\gamma: [0,\,1] \rightarrow \mathbb{R} ^n$ with $\frac{\partial \gamma}{\partial s} \neq 0 \,\, \forall s \in[0,\,1] $. Define $e(s,\gamma) = \gamma_s^T(s) M(\gamma(s)) \gamma_s(s)$ where $ \gamma_s =  \frac{\partial \gamma}{\partial s}$. Then the length of the curve with end points at $\gamma(0)$ and $\gamma(1)$ is defined as:
\begin{equation}
\ell (\gamma) = \int_0^1 \sqrt{e(s,\gamma) }\, ds  \label{eqn: length}
\end{equation}
and this can be extended to piecewise-smooth curves by summing over smooth pieces. 

A Riemannian metric defines a distance between points -- the Riemannian distance -- given by the infimum of lengths of curves joining the end points. If the resulting metric space is complete, then the Hopf-Rinow theorem states that a minimizing path exists, which is a geodesic \cite{boothby_introduction_1986}. Geodesics are analogues of straight lines in curved space. A special property of a geodesic is that it has constant speed, i.e. $e(s,\gamma)$ is independent of $s\in[0,1]$. If $\gamma$ is a geodesic, we have
\[ \ell (\gamma)^2 = \int_0^1 e(s, \gamma)\, ds  =: E(\gamma)\, ,\]
where $E(\gamma)$ is referred to as the energy of $\gamma$. So paths of minimum energy also have minimum length, and for numerical optimization the energy is preferred due to smoothness. 

Classically, a minimal geodesic would be found by using the Euler-Lagrange equation to obtain necessary conditions in the form of an ordinary differential equation \cite{Arnold}, but solving the associated two-point boundary value problem is generally non-trivial. Alternatively, the search for a geodesic can be expressed as a direct optimization over smooth paths:
\begin{equation} \label{eqn: geodesic optimisation}
\underset{\gamma \in C[0,\, 1]} {\mathrm{arg\,min}} \; \int_0^1  e(s, \gamma) \, ds\, \quad \text{s.t} \quad  \gamma(0) = x^*(t), \quad \gamma(1) = x(t)\, .
\end{equation}
In this paper, we will use a pseudospectral method to discretize this problem, and then apply a quasi-Newton algorithm to the resulting nonlinear program.

\subsection{Control Contraction Metrics}

Contraction analysis, as presented in \cite{Lohmiller}, is based upon the study of the differential (linearized) dynamics of \eqref{eqn: system dynamics}.
Roughly speaking, if all solutions of a system are locally stable, then all solutions are globally stable \cite{Lohmiller}. This can be established by way of a contraction {\em metric}, i.e. a Riemannian metric $M(x,t)$ for which the associated differential lengths $\sqrt{\delta_x^TM(x,t)\delta_x}$ shrink exponentially with time.

A CCM is a Riemannian metric for which differential lengths can be {\em made} to shrink by control action and more details about contraction theory can be found in \cite{Manchester4}. A central result in \cite{Manchester4} is that the search for a CCM, $M(x,t)$, is equivalent to finding a dual CCM $W(x,t) = M^{-1}(x,t)$. The search for $W(x,t)$ (and $\rho(x,t)$) is done by solving the following Linear Matrix Inequality (LMI): $$-\dot{W} + WA^T+ AW - \rho BB^T< -2\lambda W$$ where $A$ is the Jacobian matrix and $B$ is the control matrix.
Note that the LMI is jointly convex in $W(x,t)$ and $\rho(x,t)$. If \eqref{eqn: system dynamics} has polynomial dynamics, then sum-of-squares (SOS) programming \cite{parrilo2003semidefinite} is a computationally tractable method to find $W(x,t)$ and $\rho(x,t)$. Thus the offline computation required for a CCM controller is a convex problem and is tractable through SOS programming.

Using $W(x,t)$ and $\rho(x,t)$ satisfying the LMI, a control signal $u(t)$ that stabilizes the system to the desired trajectory $x^\star(t)$ is given by \eqref{eqn: control signal} where $u^\star(t)$ is the nominal control input.
\begin{equation}
u(t) = u^\star(t) - \frac{1}{2} \int_0^1 \rho(\gamma(s),t) B(t)^{T}M(\gamma(s),t)\frac{\partial \gamma}{\partial s} ds \label{eqn: control signal}
\end{equation}
The integral is computed over $\gamma(s)$, a geodesic with respect to the metric $M(x,t)$ connecting $x^*(t) = \gamma(0)$ to $x(t) = \gamma(1)$.

Alternatively, one can utilize the fact that the Riemannian energy to the target state forms a control Lyapunov function, and use any control technique requiring a CLF (e.g. pointwise min-norm control) \cite[Sec. IV.A]{Manchester4}. Either way, the main computational task to solve in real-time is computation of a \textit{non-trivial} minimal geodesic associated with the CCM, i.e. a solution to Problem \eqref{eqn: geodesic optimisation}. It is precisely this computation that we address in the following sections. 

\section{Pseudospectral Method}\label{sec:pseudo}
Since \eqref{eqn: geodesic optimisation} is an infinite dimensional problem over all smooth curves, the problem needs to be discretized to make it amenable to numerical solutions. Since geodesics are smooth paths on smooth manifolds, polynomials are a natural class of basis functions. In the pseudospectral context, the discretization points are called collocation points or nodes. A good choice of polynomial basis, collocation points and integration scheme are essential for successful application. 

In parametrizing the geodesic $\gamma(s)$, we will represent each state $\lbrace x_1,\,x_2,\,x_3,...,x_n \rbrace$ along the geodesic with \eqref{eqn: linear combination polynomial}. $c_{ij}$ is the $jth$ coefficient of $\gamma_i$, $D$ is the maximum degree of the polynomial and $\phi_j(s)$ is the $jth$ polynomial basis function.
\begin{equation}
 \gamma(s) = [\gamma_1(s) , \gamma_2(s),..., \gamma_n(s) ]^T \;\; , \;\; \gamma_i(s) = \sum_{j=0}^D c_{ij}\phi_j(s).\label{eqn: linear combination polynomial}
\end{equation}
The Chebyshev Pseudospectral method is a popular technique in solving optimal control problems \cite{GongFahrooRoss_PS_Cheby}. The name comes from using Chebyshev polynomials to define the nodes. In particular, the Chebyshev-Gauss-Lobatto (CGL) nodes are represented by a simple analytic formula and can be mapped to any interval through an affine transformation. Further, the Clenshaw-Curtis quadrature (CCQ) scheme is based on CGL nodes and have weights that are given by a simple analytic formula \cite{FahrooRoss, Trefethen}. The polynomial basis associated with the Chebyshev Pseudospectral method are Lagrange interpolating polynomials. In this paper however, we analyze a variation of this Chebyshev Pseudospectral method - we show that for our case study of a stiff nonlinear system, a Chebyshev polynomial basis is more effective at solving the geodesic problem than Lagrange polynomials.

\subsection{Discretizaton of the Problem} \label{sec:pseudo_formulation}
We follow the pseudospectral method outlined in \cite{GongFahrooRoss_PS_Cheby} but our problem is simpler due to the \textit{lack of dynamic constraints} and Chebyshev polynomials are used as the basis, rather than Lagrange polynomials. Here, we will describe the specifics of our formulation.

Following on from \eqref{eqn: linear combination polynomial}, we can write \eqref{eqn: parametrised x delta} where $T_k(s)$ is the $k^{th}$ degree Chebyshev polynomial of the first kind defined on the $\left[ 0,\, 1\right]$ interval. $\mathbf{c}$ is the decision vector containing all the coefficients: $$\mathbf{c} = (c_{10} ,\,  c_{11} , ...   c_{1D} ,\,  c_{20} ,\,  c_{21} ,...  c_{2D} ,\,...  c_{n0} ,\,  c_{n1} ,...  c_{nD} )\,. $$
Since the coefficients appear linearly, the differentiation needed for gradient descent can be derived analytically.
\begin{equation}
 \gamma_i(\mathbf{c};s) = \sum_{j=0}^D c_{ij}T_j(s), \quad \frac{\partial \gamma_i}{\partial s}:=\gamma_{s_i}(\mathbf{c};s) = \sum_{j=0}^D c_{ij}\frac{d T_k(s)}{ds} \label{eqn: parametrised x delta}
\end{equation}
Recall $D$ is the degree of the polynomial and $n$ is the dimension of the state space in \eqref{eqn: system dynamics}. So the number of unknowns for the problem will be $(D+1)\times n$. We will define $N$ as the stopping index for our CGL nodes $\lbrace{s_0, s_1, ..., s_N\rbrace}$.
Integration of a function along a curve is approximated using the CCQ quadrature scheme evaluated at the CGL nodes. We refer the reader to \cite{GongFahrooRoss_PS_Cheby} for the formulation. Essentially, we can evaluate \eqref{eqn: geodesic optimisation} for our CCM using this scheme where $w_k$ are the CCQ weights.
\begin{equation} \label{CCQ rule}
\int_0^1 \gamma_s^T M(\gamma,t)\gamma_s ds = \sum_{k=0}^N	\gamma_s(s_k)^T M(\gamma(s_k),t)\gamma_s(s_k)w_k 
\end{equation}
The CCQ representation shown in \eqref{CCQ rule} is exact when $N=D$ \emph{and} if the integrand is polynomial. However, we cannot guarantee that $M(x,t)$ is polynomial, even though $W(x,t)$ is solved through SOS programming. As such, we require $N > D$ to improve the accuracy of the quadrature scheme. This is the primary motivation behind selecting a Chebyshev polynomial basis over Lagrange polynomials. Lagrange polynomials are defined by the number of nodes, meaning that we always have $D=N$. Hence we cannot guarantee accuracy for our quadrature scheme and as we increase the number of nodes, the size of our problem grows. While Chebyshev polynomials are defined independent of $N$. An exploration for a sufficiently large $N$ for our particular stiff system is given in Section~\ref{sec:finding_a_geodesic}.
Chebyshev polynomials also give rise to a simple endpoint constraint since $T_k(0) = 0$ if $k$ is even or $T_k(0) = -1$ if $k$ is odd, and $T_k(1) = 1$.

We can now substitute \eqref{eqn: parametrised x delta} into our optimization problem \eqref{eqn: geodesic optimisation} and see that our constraints \eqref{eqn: linear constraints matrix} are now linear. Here, $P = [T_0(0), T_1(0) ,T_2(0), ... , T_D(0)]$ is a row vector and $\mathbf{1}$ a row vector of ones of the same length. $BDiag_n(F)$ is a block diagonal matrix with the matrix $F$ down the diagonal $n$ times.
 
\begin{equation}
\mathcal{C}(\mathbf{c}) :=\underbrace{ 
\begin{bmatrix}
BDiag_n(P)\\
BDiag_n(\mathbf{1})
\end{bmatrix}}_{A_c} \mathbf{c} - \underbrace{\begin{bmatrix}
x^*(t) \\
x(t)
\end{bmatrix}}_{b_c} = 0  \label{eqn: linear constraints matrix}
\end{equation}

As such, the optimization problem in \eqref{eqn: geodesic optimisation} can be expressed by \eqref{eqn: Chebyshev optimisation problem}. This is a smooth nonlinear optimization problem with linear constraints and $\mathbf{c}$ is the optimization variable. We will refer to objective function as $E(\mathbf{c})$ in future sections.
\begin{equation} \label{eqn: Chebyshev optimisation problem}
\begin{split}
\underset{\mathbf{c} \in \mathbb{R}^{n(D+1)}} {\mathrm{argmin}} \quad &  \sum_{k=0}^N  \gamma_s(\mathbf{c},s_k)^T M(\gamma(\mathbf{c},s_k),t) \gamma_s(\mathbf{c},s_k) \, w_k \\
\text{s.t} \qquad & \mathcal{C}(\mathbf{c}) = 0
\end{split}
\end{equation}

\subsection{Quasi-Newton Optimization} 
This section will describe important aspects of the optimization procedure used to solve \eqref{eqn: Chebyshev optimisation problem}, in particular, give an analytic representation of the gradient. 
To improve computation costs we employ the standard BFGS quasi-Newton strategy to approximate the Hessian using the gradient. To ensure first-order optimality conditions are met, the KKT matrix is solved at each iteration of the optimization algorithm \cite{Nocedal}.
Every component in the KKT matrix is readily known or can be computed with ease; $A_c$ and $b_c$ are described in \eqref{eqn: linear constraints matrix}, $H$ is approximated using BFGS and $g$ can be calculated analytically. We use the relation $\dot{M} = -M\dot{W}M$ (the dot represents a differentiation in any argument). Thus each component of $g$ is given by \eqref{eqn: gradient}. (The arguments are omitted for a concise representation.)
\begin{equation} \label{eqn: gradient}
\frac{\partial E}{\partial c_{ij}} = \sum_{k=1}^N\, 2\gamma_s^TM\,\frac{d T_k}{ds}w_k-\gamma_s^TM\frac{\partial W}{\partial x_i}M \gamma_s T_j w_k  \Big|_{\mathbf{c}, s_k}
\end{equation}

The initialization for the optimization variable $\mathbf{c}$ is simply a straight line connecting $x^*(t)$ to $x(t)$. 

Hyperparameters of the quasi-Newton descent algorithm include the tolerance $\beta= 10^{-10}$ to terminate the descent, initial step size of the line search $\alpha_0=1$, termination condition for the line search $\bar{c}=0.1$ and the rescaling factor $\tau=0.1$ for the backtracking line search. However, these parameters are not too expensive, and are set at typical values to ensure reliable convergence.

The expensive parameters are the degree of polynomial used and the number of nodes. The minimal degree and number of nodes depend on the CCM and the start and end points. If the CCM is constant, then geodesics are trivially straight lines. Alternatively, if the CCM is state dependent then geodesic will tend to be nonlinear and hence a higher degree and more nodes would be needed.

\subsection{Adaptive Degree Selection} \label{subsec: Finding the Geodesic}
In this section, we discuss a method in selecting the minimal $D$ and $N$ to obtain a geodesic with sufficient accuracy. Recall that a geodesic has constant energy along the path. Since this constraint is not enforced during the optimization, the deviation of $e(s, \gamma)$ from the constant energy can be used as an independent validation that the solution found is a geodesic. The constant energy constraint cannot be enforced beforehand because the energy is not known a priori. When our optimization algorithm converges to the solution $\mathbf{c}^*$, let $E(\mathbf{c}^*) := E^*$ be the optimal energy.

Then to measure the difference between the energy along the geodesics and $E^*$ we use \eqref{eqn: area between energy curve} as a measure of accuracy. This can be interpreted as relative root-squared-error because we want to take into account the increases in nonlinearity as the metric is more nonlinear, or the points are further apart (i.e., $E^*$ is larger). 

\begin{equation} \label{eqn: area between energy curve}
\mathcal{E} = \frac{1}{E^*} \sqrt{\int_0^1 \left( e(s, \gamma) - E^* \right)^2  ds}
\end{equation}

For a true geodesic $\mathcal{E}=0$. Given the facts: (i) set of all polynomials on $\left[ 0,\, 1\right]$ spans the set of smooth curves, (ii) by increasing the maximum degree we enlarge the set of curves that can be represented, and (iii) the curve of minimal energy has constant speed, we propose to simply increase the degree until  $\mathcal{E}$ is reduced below a certain threshold. It was found that $\mathcal{E} < 10^{-6}$ was a sufficient tolerance.

As mentioned before, we require $D<N$ in our optimization. Since we also want to keep $N$ small to reduce computation costs, we consider values for $N$ in the form of $N = D + a$. This offers smallest possible $N$ values without sacrificing too much on computation time. Section \ref{subsec: chebyshev pseudospectral method} will explore a suitable value for $a$.

\section{Illustrative Example}\label{sec:exGeo}
To illustrate how a geodesic is computed and used to construct a CCM controller, the system given in \eqref{eqn: Andrieu system} is used as the subject of this case study. This system was investigated in \cite{Manchester4} and \cite{Andrieu}; it is an interesting system because it is not feedback linearizable, local controllers fail globally and the dynamics are stiff and highly unstable. In \cite{Andrieu}, the problem of uniting a locally optimal and globally stabilizing controller is explored. In the CCM framework, this problem can be easily formulated because the search for $W(x,t)$ and $\rho(x,t)$ is over a convex set.
\begin{equation} \label{eqn: Andrieu system}
\begin{bmatrix}
\dot{x}_1 \\
\dot{x}_2 \\
\dot{x}_3 
\end{bmatrix}  = \begin{bmatrix}
-x_1 + x_3\\
x_1^2-x_2-2x_1x_3+x_3\\
-x_2
\end{bmatrix} + \begin{bmatrix}
0\\0\\1
\end{bmatrix}u{}
\end{equation}
Sum-of-squares programming was carried out on Yalmip \cite{Yalmip} and the semidefinite programming solver MOSEK. The metric is enforced to match LQR locally about zero \cite{Manchester4} by having $W(0,t) = P^{-1}$ where $P$ is the solution to the algebraic Riccati equation (with $Q=R=I$). A CCM was found and it is of the form $ W(x,t) = W_0 + W_1 x_1 + W_2 x_1^2 $ where $W_i$ are constant matrices and $ \rho(x,t) = \rho_0 + \rho_1x_1 + \rho_2x_1^2 $. Note that $W$ and $\rho$ are quadratic in one variable, which is also the quadratic term in \eqref{eqn: Andrieu system}. It was shown in \cite{Manchester4} that metrics of this form are complete, and thus a minimal geodesic exists between every pair of points. 
\section{Controller Design}\label{sec:exCon}
In this section, we compare the performance of using a CCM controller to two popular methods for feedback control: LQR and NMPC. A CCM controller is constructed at each time step by solving the geodesic problem using the proposed pseudospectral method and integrating the differential controller along this path. The LQR controller was linearized at the origin while a multiple-shooting NMPC regime was implemented on ACADO \cite{Houska2011a}, an open-source C++ software that solves optimal control problems. ACADO approximates $u$ as piecewise constant, and the state trajectories as piecewise linear.
\subsection{Starting Far From the Origin}
\begin{figure}[t]
	\centering
	\includegraphics[width=0.45\textwidth]{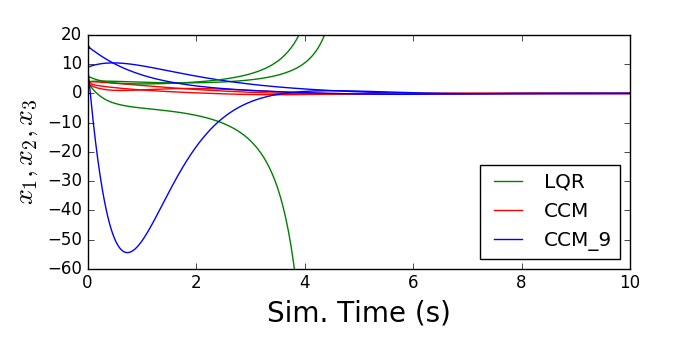}
	\caption{States starting at $[4,4,6]^T$ using LQR and CCM controllers (green and red). States starting at $[9,9,9]^T$ (blue) is stabilized using a CCM controller.}
	\label{fig: CCM_LQR_446}
\end{figure}

First we look at cases where the initial condition is relatively far away from LQR's point of linearization to investigate the region of stability each controller exhibits. When starting at $[4,4,6]^T$ and the target state $x^*(t) = [0,0,0]^T$, the LQR controller fails to stabilize the system, even though it is successful at other nearby points, e.g., $[5,5,5]^T$. This highlights the weakness in global stability of LQR. On the other hand, a CCM controller was successful in stabilizing the system and Figure \ref{fig: CCM_LQR_446} illustrates this. Another starting point further out at $[9,9,9]^T$ (CCM\_9 in Figure \ref{fig: CCM_LQR_446}) was also tested and the CCM was still able to stabilize the system. 
The stiffness of the system is evident, indicated by the very rapid and slow dynamics. ACADO was not able to compute a solution from this initial condition, which we believe is due to the stiffness of the ODE: the slow dynamics require a long time horizon, and the fast dynamics require a short sampling time, leading to a very large number of required nodes over a long time horizon.
\subsection{Starting Close to the Origin}
\begin{figure}[t]
\centering
\includegraphics[width=0.45\textwidth]{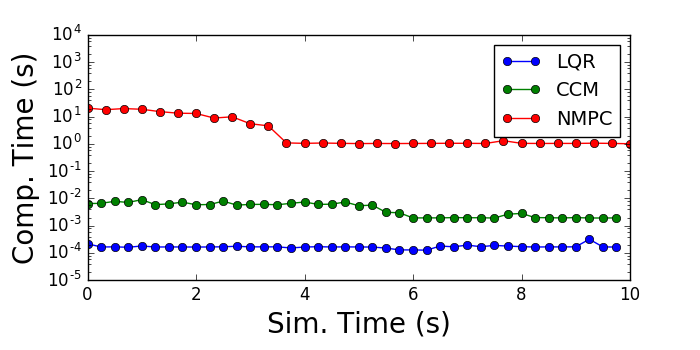}
\caption{Comparison of average computation time of a LQR, CCM and NMPC controller at each iteration during a simulation with states starting at $[1,1,1]^T$.}
\label{fig: 111 computation time}
\end{figure}

On the other hand, considering initial states close to the origin where the stiffness is not as prevalent, we find that all CCM, LQR and NMPC controllers produce very similar results. For an initial state of $x(0) = [1,1,1]^T$ and target state of $x^*(t) = [0,0,0]^T$, the state trajectories are almost indistinguishable since they are all locally optimal in an LQ sense. Clearly LQR is superior in terms of computation time since it only uses elementary matrix operations and it is often the technique used when operating close to the point of linearization. However, Figure \ref{fig: 111 computation time} compares the computation time between LQR, CCM and NMPC. We see that the computation of CCM is reasonably low and with the advantage that is can also stabilize far from the origin. As such, CCM poses as a nice middle ground in terms of computational costs and global stability.
To summarize: for this system, LQR was effective only for regions local to the point of linearization, while multiple shooting NMPC using ACADO required many discretization points and did not succeed for larger initial conditions. CCM gave stabilizing results with starting points far from and near the origin. The following section will discuss ways the geodesic can be computed in order to rapidly construct the CCM controller.

\section{Finding a Geodesic} 
\label{sec:finding_a_geodesic}
\begin{figure*}[t]
 	\centering
 	\includegraphics[width=\textwidth]{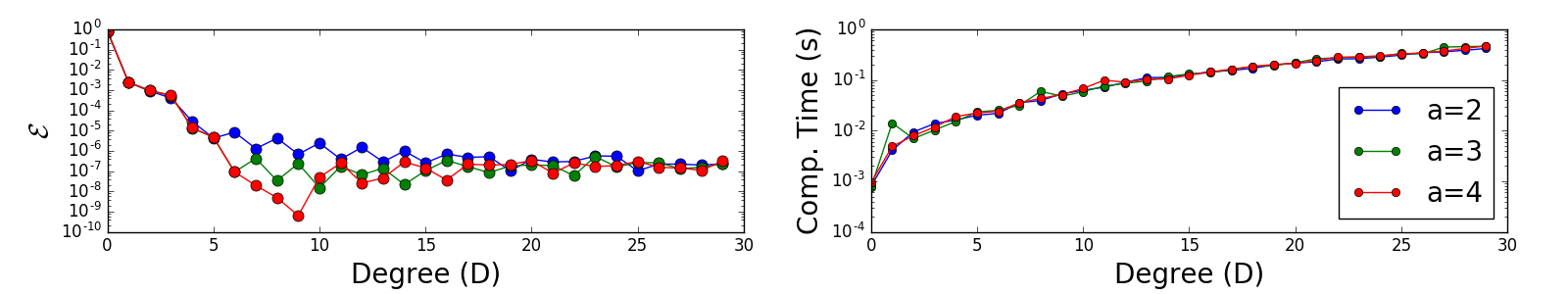}
 	\caption{Left: $\mathcal{E}$ versus $D$ with various values of $a$. End points are $x(t) = [9,9,9]^T$ and $x^*(t)=[0,0,0]^T$. Right: Average computation time versus $D$ for same end points.}
 	\label{fig: error_vs_K_add}
 \end{figure*}
We have seen that using a CCM controller was effective in stabilizing \eqref{eqn: Andrieu system} while LQR and NMPC was unsuccessful. This section will two explore ways to compute the geodesics which allows the construction of a CCM controller possible. As such, we shall see that the proposed method, as outlined in Section \ref{sec:pseudo_formulation}, is very fast and accurate, making it possible for real-time applications.
\subsection{Pseudospectral Method}
\label{subsec: chebyshev pseudospectral method}
First, we numerically validate our choice of Chebyshev polynomials over Lagrange interpolating polynomials. For relatively close end points that do not require high degrees, Chebyshev polynomials yield faster computation times than using Lagrange polynomials. For example, computation time for Lagrange polynomials was on the order of $10^{-2}$ while it was on the order of $10^{-3}$ for Chebyshev polynomials (with endpoints, $[0, 0, 0]^T$ and $[1, 1, 1]^T$). This implies that Lagrange polynomials will perform even worse for higher degrees/nodes As such, since a Chebyshev polynomial basis offer more rapid convergence, this is our choice of basis.

First, a value for $a$ must be fixed. With starting states at $[9,9,9]^T$, the effect on $\mathcal{E}$ was tested with different values of $a$ and this is given in Figure \ref{fig: error_vs_K_add} (left). $a=4$ was chosen because it gave results that were below the tolerance ($10^{-6}$) for low degrees without a huge cost in computation time (Figure \ref{fig: error_vs_K_add} right). The results were implemented in Julia, running on a 2012 MacBook Pro with a 2.90GHz Intel Core i7 processor. The code is available at \verb|https://github.com/karenl7/GeodesicCCM|. 

With $a=4$ fixed, the degree, $\mathcal{E}$ and the average computation time for various initialization points are given in Table \ref{tab: Geodesics for different end point}. It can be seen that only a degree 7 polynomial was needed for a far away starting point at $[9,9,9]^T$. As a point of comparison for the same end points $N=D=70$ was needed for Lagrange polynomials to achieve acceptable accuracy, at an average computation time of 19.7 seconds. Hence we choose Chebyshev polynomials over Lagrange polynomials.

\begin{table}[t]
\centering
\caption{Average computation time to compute a geodesic with different starting points and $x^*(t) = 0$.}
\label{tab: Geodesics for different end point}
\begin{tabular}{|c|c|c|}
\hline
\bf $x(t)$ & \bf  $D$ & \bf Avg. comp. time \\ \hline
$[1,1,1]^T$ & 4 & 0.00811\\
$[3,3,3]^T$ & 4 & 0.00848\\
$[5,5,5]^T$ & 5 & 0.01487\\ 
$[7,7,7]^T$ & 6 & 0.02069\\ 
$[9,9,9]^T$ & 7 & 0.02542\\ \hline
\end{tabular}
\end{table} 
Since computation time depends on the degree and number of nodes used, possible offline computations to speed up the online computation could involve pre-computing optimal $D$ and $a$ values for regions in the state space. These values can be accessed, rather than found adaptively during real-time implementation. However, this could get very complex for higher dimensional problems and when $x^*(t)$ is not constant.

Nonetheless, the proposed Chebyshev pseudospectral method with Chebyshev polynomials as basis functions offer a very quick and efficient way of approximating a geodesic with a sufficient accuracy.
\subsection{Multiple Shooting using ACADO}
The geodesic problem can also be expressed as a simple OCP by introducing trivial dynamical constraints and this is given by $\dot{x} = u  , \quad J = \int_0^1 u^TM(x)u\, dt$. The computation time to find the geodesic using ACADO was significantly greater than the Chebyshev pseudospectral method. Even converging to a solution for a path connecting $[1,1,1]^T$ to $[0,0,0]^T$ took multiple orders of magnitude greater than the pseudospectral method and with significantly less accuracy. 
For example, with 100 segments, it took roughly 8.8 seconds and $\mathcal{E} > 10^{-2}$. Increasing the number of segments would certainly improve the accuracy, but this would severely increase computation time, and would be impractical for real-time implementation. 
This is perhaps due to the need to invert $W(x,t)$ to obtain $M(x,t)$ and take derivatives in the ACADO toolkit which in general is expensive. However, with the pseudospectral framework, we could utilize the formula for the gradient.
As such, compared to the multiple shooting method in ACADO, the proposed pseudospectral method is significantly faster and more accurate even with less nodes. 
\section{CONCLUSIONS}
We examined the online computation required for a CCM controller. The benefits associated with a CCM controller include tractable offline computations, guarantees on global stability and, as presented in this paper, rapid and tractable online computations. It was shown that for a particular stiff system, a CCM controller was able to stabilize the system beyond LQR's region of stability and that the stiffness made multiple-shooting for NMPC very inefficient. 
Thus CCM has the potential to compute nonlinear stabilizing controllers in real-time for a larger range of initialization points. 
A pseudospectral method using Chebyshev polynomials to solve the geodesic problem was found to be fast and tractable, amenable for real-time implementation. 
We conclude that the CCM/pseudospectral method offers a viable alternative for certain difficult nonlinear stabilization problems, since it encompasses the simplicity of LQR and the global performance of NMPC.

\addtolength{\textheight}{0.0cm} 


\bibliography{ACCKaren.bib}

\begin{thebibliography}{10}
\providecommand{\url}[1]{#1}
\csname url@rmstyle\endcsname
\providecommand{\newblock}{\relax}
\providecommand{\bibinfo}[2]{#2}
\providecommand\BIBentrySTDinterwordspacing{\spaceskip=0pt\relax}
\providecommand\BIBentryALTinterwordstretchfactor{4}
\providecommand\BIBentryALTinterwordspacing{\spaceskip=\fontdimen2\font plus
\BIBentryALTinterwordstretchfactor\fontdimen3\font minus
  \fontdimen4\font\relax}
\providecommand\BIBforeignlanguage[2]{{%
\expandafter\ifx\csname l@#1\endcsname\relax
\typeout{** WARNING: IEEEtran.bst: No hyphenation pattern has been}%
\typeout{** loaded for the language `#1'. Using the pattern for}%
\typeout{** the default language instead.}%
\else
\language=\csname l@#1\endcsname
\fi
#2}}

\bibitem{Allgower}
F.~Allg{\"o}wer and A.~Zheng, \emph{Nonlinear model predictive control}.\hskip
  1em plus 0.5em minus 0.4em\relax Birkhauser, 2012, vol.~26.

\bibitem{sontag_universal_1989}
E.~D. Sontag, ``A ‘universal’ construction of {Artstein}'s theorem on
  nonlinear stabilization,'' \emph{Systems \& Control Letters}, vol.~13, no.~2,
  pp. 117--123, Aug. 1989.

\bibitem{Krstic}
M.~Krstic, I.~Kanellakopoulos, and P.~Kokotovic, \emph{Nonlinear and adaptive
  control design}.\hskip 1em plus 0.5em minus 0.4em\relax Wiley, 1995, vol.
  222.

\bibitem{Lohmiller}
W.~Lohmiller and J.-J.~E. Slotine, ``On {Contraction} {Analysis} for
  {Non}-linear {Systems},'' \emph{Automatica}, vol.~34, no.~6, pp. 683--696,
  June 1998.

\bibitem{CCM}
I.~R. Manchester and J.-J.~E. Slotine, ``Control {Contraction} {Metrics} and
  {Universal} {Stabilizability},'' in \emph{Proceedings of the {IFAC} {World}
  {Congress}}, Cape Town, South Africa, 2014.

\bibitem{Manchester4}
------, ``Control contraction metrics: Convex and intrinsic criteria for
  nonlinear feedback design,'' \emph{IEEE Transactions on Automatic Control},
  In Press 2017.

\bibitem{Aylward}
E.~M. Aylward, P.~A. Parrilo, and J.~E. Slotine, ``Stability and robustness
  analysis of nonlinear systems via contraction metrics and {SOS}
  programming,'' \emph{Automatica}, vol.~44, no.~8, pp. 2163--2170, 2008.

\bibitem{ManchesterTang}
I.~R. Manchester, J.~Z. Tang, and J.~E. Slotine, ``Unifying classical and
  optimization-based methods for robot tracking control with control
  contraction metrics,'' \emph{International Symposium on Robotics Research
  (ISRR)}, pp. 1 -- 16, 2015.

\bibitem{Arnold}
V.~I. Arnold, \emph{Mathematical methods of classical mechanics}.\hskip 1em
  plus 0.5em minus 0.4em\relax Springer Science \& Business Media, 1989,
  vol.~60.

\bibitem{Ying}
L.~Ying and E.~J. Candes, ``Fast geodesics computation with the phase flow
  method,'' \emph{Journal of computational physics}, vol. 220, no.~1, pp.
  6--18, 2006.

\bibitem{Kimmel}
R.~Kimmel and J.~A. Sethian, ``Computing geodesic paths on manifolds,''
  \emph{Proceedings of the National Academy of Sciences}, vol.~95, no.~15, pp.
  8431--8435, 1998.

\bibitem{boykov_computing_2003}
Y.~Boykov and V.~Kolmogorov, ``Computing geodesics and minimal surfaces via
  graph cuts,'' in \emph{Proc. {IEEE} {International} {Conference} on
  {Computer} {Vision}}, 2003.

\bibitem{diehl_efficient_2009}
M.~Diehl, H.~J. Ferreau, and N.~Haverbeke, ``Efficient numerical methods for
  nonlinear {MPC} and moving horizon estimation,'' in \emph{Nonlinear {Model}
  {Predictive} {Control}}.\hskip 1em plus 0.5em minus 0.4em\relax Springer,
  2009, pp. 391--417.

\bibitem{benson_direct_2006}
D.~A. Benson, G.~T. Huntington, T.~P. Thorvaldsen, and A.~V. Rao, ``Direct
  trajectory optimization and costate estimation via an orthogonal collocation
  method,'' \emph{Journal of Guidance, Control, and Dynamics}, vol.~29, no.~6,
  pp. 1435--1440, 2006.

\bibitem{Elnagar}
G.~N. Elnagar and M.~A. Kazemi, ``Pseudospectral {Chebyshev} optimal control of
  constrained nonlinear dynamical systems,'' \emph{Computational Optimization
  and Applications}, vol.~11, no.~2, pp. 195--217, 1998.

\bibitem{FahrooRoss}
F.~Fahroo and I.~M. Ross, ``Direct trajectory optimization by a {Chebyshev}
  pseudospectral method,'' \emph{Journal of Guidance, Control, and Dynamics},
  vol.~25, no.~1, pp. 160--166, 2002.

\bibitem{Garg}
D.~Garg, M.~Patterson, W.~W. Hager, A.~V. Rao, D.~A. Benson, and G.~T.
  Huntington, ``A unified framework for the numerical solution of optimal
  control problems using pseudospectral methods,'' \emph{Automatica}, vol.~46,
  no.~11, pp. 1843--1851, 2010.

\bibitem{Williams}
P.~Williams, ``Application of pseudospectral methods for receding horizon
  control,'' \emph{Journal of Guidance, Control, and Dynamics}, vol.~27, no.~2,
  pp. 310--314, 2004.

\bibitem{FahrooRoss_PS_infRHC}
F.~Fahroo and I.~M. Ross, ``Pseudospectral methods for infinite-horizon
  nonlinear optimal control problems,'' \emph{Journal of Guidance, Control, and
  Dynamics}, vol.~31, no.~4, pp. 927--936, 2008.

\bibitem{GongFahrooRoss_PS_Cheby}
Q.~Gong, I.~M. Ross, and F.~Fahroo, ``A chebyshev pseudospectral method for
  nonlinear constrained optimal control problems,'' \emph{Joint 48th IEEE
  Conference on Decision and Control and 28th Chinese Control Conference},
  2009.

\bibitem{ross2012review}
I.~M. Ross and M.~Karpenko, ``A review of pseudospectral optimal control: from
  theory to flight,'' \emph{Annual Reviews in Control}, vol.~36, no.~2, pp.
  182--197, 2012.

\bibitem{Lee}
J.~M. Lee, \emph{Riemannian manifolds: an introduction to curvature}.\hskip 1em
  plus 0.5em minus 0.4em\relax Springer Science \& Business Media, 2006, vol.
  176.

\bibitem{boothby_introduction_1986}
W.~M. Boothby, \emph{An introduction to differentiable manifolds and
  {Riemannian} geometry}.\hskip 1em plus 0.5em minus 0.4em\relax Academic
  press, 1986.

\bibitem{parrilo2003semidefinite}
P.~A. Parrilo, ``Semidefinite programming relaxations for semialgebraic
  problems,'' \emph{Mathematical programming}, vol.~96, no.~2, pp. 293--320,
  2003.

\bibitem{Trefethen}
L.~N. Trefethen, ``Is {Gauss} quadrature better than {Clenshaw}-{Curtis}?''
  \emph{SIAM review}, vol.~50, no.~1, pp. 67--87, 2008.

\bibitem{Nocedal}
J.~Nocedal and S.~Wright, \emph{Numerical optimization}.\hskip 1em plus 0.5em
  minus 0.4em\relax Springer Science \& Business Media, 2006.

\bibitem{Andrieu}
V.~Andrieu and C.~Prieur, ``Uniting two control lyapunov functions for affine
  systems,'' \emph{IEEE Transactions on Automatic Control}, vol.~55, no.~8, pp.
  1923--1927, 2010.

\bibitem{Yalmip}
J.~L{\"o}fberg, ``Yalmip: A toolbox for modeling and optimization in matlab,''
  in \emph{Computer Aided Control Systems Design, 2004 IEEE International
  Symposium on}.\hskip 1em plus 0.5em minus 0.4em\relax IEEE, 2004, pp.
  284--289.

\bibitem{Houska2011a}
B.~Houska, H.~J. Ferreau, and M.~Diehl, ``\BIBforeignlanguage{en}{{ACADO}
  toolkit-{An} open-source framework for automatic control and dynamic
  optimization},'' \emph{\BIBforeignlanguage{en}{Optimal Control Applications
  and Methods}}, vol.~32, no.~3, pp. 298--312, May 2011.

\end{thebibliography}

\end{document}